\documentclass[aps,jap,twocolumn,superscriptaddress,showpacs]{revtex4}

\usepackage{graphicx}
\usepackage{dcolumn}
\usepackage{bm}

\begin{document}

\preprint{APS/123-QED}

\title{Ion beam sputtering method for progressive\\ reduction of nanostructures dimensions}

\author{M. Savolainen}
\email{marko.savolainen@phys.jyu.fi}
\author{V. Touboltsev}
\author{P. Koppinen}
\author{K.-P. Riikonen}
\author{K. Arutyunov} 
\affiliation{%
NanoScience Center, Department of Physics, University of Jyv\"askyl\"a
\\
PB 35 (YFL), FIN-40014 University of Jyv\"askyl\"a\\
Jyv\"askyl\"a, Finland
}%

\date{\today}

\begin{abstract}
An ion beam based dry etching method has been developed for progressive reduction of dimensions of 
prefabricated nanostructures. The method has been successfully applied to aluminum nanowires and 
aluminum single electron transistors (SET). The method is based on removal of material from the 
structures when exposed to energetic argon ions and it was shown to be applicable multiple times 
to the same sample. The electrical measurements and samples imaging in between the sputtering 
sessions clearly indicated that the dimensions, i.e. cross-section of the nanowires and area of 
the tunnel junctions in SET, were progressively reduced without noticeable degradation of the 
sample structure. We were able to reduce the effective diameter of aluminum nanowires from 
$\sim$65 nm down to $\sim$30 nm, whereas the tunnel junction area has been reduced by 40 \%.
\end{abstract}

\maketitle

\section{Introduction}
There is a variety of different techniques available for fabrication of nano- or micron-sized 
structures. Ultraviolet (UV) lithography is widely used in microelectronic industry to fabricate 
large-scale integrated circuits with vast amount of functional elements at once. However, the 
minimum lateral dimensions attainable with this technique is about 250 nm. By using deep UV light 
the limit might be pushed close to 100 nm in the future\cite{xia}. More advanced methods based on 
electron beam lithography (EBL) are capable to provide even smaller dimensions and have been 
applied, e.g., for fabrication of 5-7 nm wide etched lines on a silicon substrate\cite{ChenAhmed}. 
However, when evaporating metallic structures through masks made with EBL the limit is higher, 
around 20-50 nm, depending on the molecule size of the resist material and the performance of the 
particular equipment. The disadvantage of EBL is that it is rather slow. X-rays lithography can in 
principle be used for patterning, but this method requires significantly more efforts and 
complicated masks\cite{xia}. It is also possible to use the sharp tip of an atomic force 
microscope (AFM) to transfer single particles on a substrate to form a nanopattern \cite{lindeli}. 
Alternatively, one can oxidize patterns on the hydrogen-passivated surface with the tip or scratch 
the pattern on a thin resist layer (see Ref.~\onlinecite{Anssi} and references therein for 
detailed description of the AFM based methods). The difficulty in using AFM in patterning is in 
removal of the mask material (lift-off) after metal evaporation. As a result, at present moment 
AFM based nanofabrication has rather limited range of applications.

We studied a different kind of approach where the dimensions of the prefabricated nanosized 
structures are reduced by ion beam sputtering in controllable and reproducible way. We used 
aluminum nanowires and single electron transistors (SET) to test applicability of the sputtering 
method. SET is one of the fundamental components in nanoelectronics and ultrasmall tunnel 
junctions in general have a lot of potential applications in the future\cite{pashkin}. The change 
in the dimensions of the tunnel junctions was detected by electrical measurements of the charging 
energy $E_{\rm C}$ at liquid helium temperature 4.2 K. In case of a nanowire, the decrease of the 
diameter was determined from the width of superconducting transition $R(T)$.  

Generally speaking, it is common to study particular properties of the system of interest as a 
function of some characteristic dimension. Traditionally, many samples of different sizes are 
fabricated for this purpose. By using the sputtering method, the electrical measurements can be 
performed on the same sample, which dimensions are progressively reduced between the measurements. 
This way the inner structure of the system stays the same and thus there are less possibilities of 
having statistical errors due to circumstantial factors in fabrication.

\section{Experimental}

\subsection{Sample fabrication}

\begin{figure}[h]
\includegraphics[width=60mm]{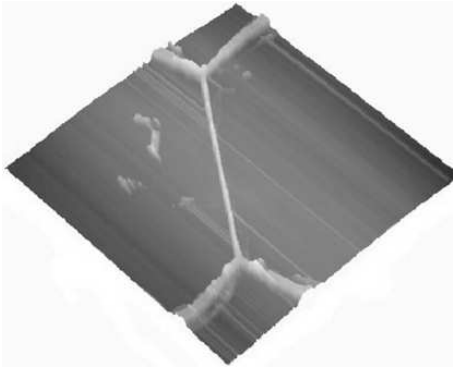}
\caption{An overview AFM-image of approximately 60 nm wide and 10 $\mu$m long aluminum nanowire.} 
\label{nanonano}
\end{figure}

All samples were fabricated on oxidized silicon substrates. Conventional EBL technique was used in 
patterning followed by metallization in an UHV (Ultra High Vacuum) chamber. Double layer PMMA/MAA 
resist was used to form an appropriate undercut structure for the angle evaporation. Nanowires 
were formed by evaporating 45 nm of aluminum on top of the substrate through the PMMA mask. Widths 
of the fabricated nanowires were approximately 50-80 nm. Figure \ref{nanonano} shows an AFM image 
of a typical sample.  

\begin{figure}[h] 
\includegraphics[width=80mm]{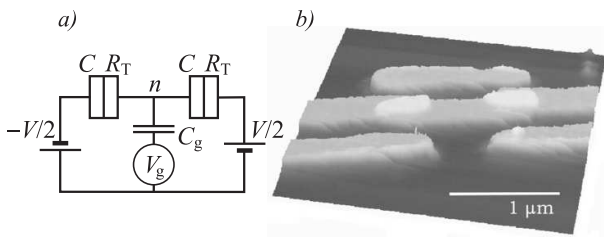}
\caption{a) Schematics of a single electron transistor biased to voltage $V$, $V_g$ is gate 
voltage. b) AFM image of the SET after ion beam etching. This image does not show the gate 
electrode. The top most and the bottom most lines are parasitic structures due to two-angle 
evaporation method.} 
\label{Setit}
\end{figure}

SETs were fabricated with a standard shadow evaporation technique. First 45 nm thick layer of 
aluminum was evaporated. Aluminum oxide barrier was grown {\it in situ} by natural oxidation in 
pure oxygen atmosphere ($\sim$20 mbar) in the loading chamber of the UHV system. After oxidation 
another 45 nm layer of aluminum was deposited from another angle on top of the previously grown 
oxide layer to form tunnel junctions.       
Fig. \ref{Setit} b) shows an AFM image of a typical SET with two Al-AlO$_x$-Al junctions. About 1 
nm thick aluminum oxide layer between the aluminum electrodes forms a tunnel junction, which is 
thin enough to provide quantum mechanical tunneling of electrons through it\cite{SET}.

\subsection{Description of the etching method}

The samples were three-dimensionally dry etched by ion beam sputtering in a set-up consisting of a 
high vacuum (p$\sim 10^{-5}$ mbar) experimental chamber equipped with a sample manipulator and 
TECTRA Electron Cyclotron Resonance (ECR) plasma ion source capable of delivering high current, 
wide and homogeneous ion beams. Before sputtering, the samples were cleaned with acetone in an 
ultrasonic bath and subsequently rinsed in isopropanol. Prior to sputtering the surface of the 
structures was always checked by profiler Tencor P15 that is capable to provide a vertical step 
height repeatability of $\sim$6-7 \AA.  For dry etching, the samples were bombarded by 1 keV 
Ar$^+$ ions to a certain fluence using an ion beam current density of about 0.014 mA/cm$^2$. In 
order to ensure uniform etching over the whole sputtered area, the ions incidence was 60$^\circ$ 
off the surface normal, and the samples were rotated while sputtering. To avoid overheating of the 
samples exposed to high current ion beam, the sample holder made of copper was cooled with water 
and the temperature while sputtering was estimated to be close to room temperature. Each sample 
contained co-evaporated strips, which were partly protected from the ion beam exposure by a 
droplet of a varnish. After etching the varnish was removed enabling the profilometer control of 
the surface step between the etched and the non-etched area. Sputtering rates for various 
materials are listed in the Table~\ref{rates}. 

\begin{table}
\caption{\label{rates}Sputtering rates of various materials [nm/min] by 1 keV Ar$^{+}$ ions (0.014 
mA/cm$^{2}$).}
\begin{ruledtabular}
\begin{tabular}{lcr}
Aluminum\footnote{The aluminum had a natural oxide layer of about 1-2 nm on the surface when 
sputtered.}&SiO$_x$/Si&Bulk Sapphire\\
\hline
1.1 & 3.9 & 0.75\\
\end{tabular}
\end{ruledtabular}
\end{table}

Ar$^+$ ions of energy 1 keV impinging on the surface at the angle of 60 degrees with respect to 
the surface normal practically do not penetrate into the subsurface layers. Penetration depth of 
the Ar$^+$ ions into Al matrix calculated by SRIM (Stopping and Ranges of Ions Matter) 
program\cite{SRIM} for the selected irradiation conditions is less than 15 \AA. Taking into 
account the high rate of the surface sputtering due to high density of the ion current and the 
glancing incidence, the ion beam etching can be considered as 'a gentle cut' of the up-most 
surface atoms without appreciable influence on the underlying layers. The surface of the samples 
was controlled before and after the ion bombardment with AFM, SEM and profilometer. The influence 
of the sputtering has a polishing effect, causing no noticeable destruction to the nanostructures.

\subsection{Electrical measurements}

The superconducting transition of the wires was measured before and after the sputtering sessions. 
The resistance of the wires as a function of temperature was measured with the four-probe method. 
The samples were inserted into a directly pumped $^4$He bath, where the temperature can be tuned 
with the accuracy of $\pm$0.1 mK.

In case of SETs the differential conductance vs. bias voltage characteristics was measured at low 
voltages ($\pm$ 10 mV per junction) at liquid helium temperature 4.2 K. The conductance 
measurements were performed with {\it Nanoway CBT Monitor 400R}, an instrument based on an AC 
resistance bridge\cite{nanoway}.

\section{Results and discussion} 

\subsection{Sputtered aluminum nanowires}

\begin{figure}[t]  
\includegraphics[width=90mm]{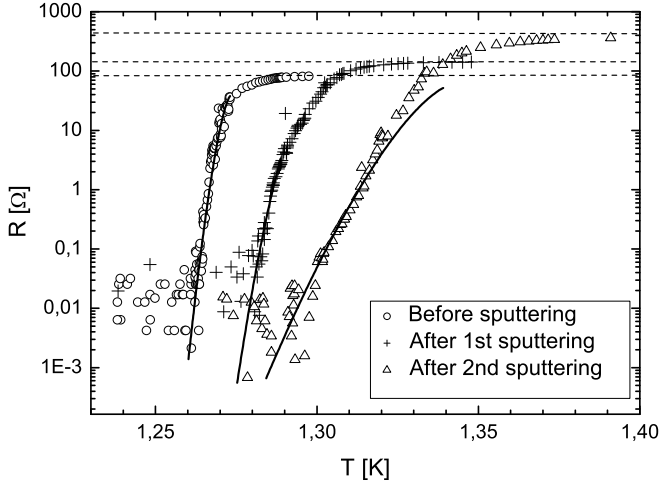}
\caption{$R(T)$ dependency of an aluminum nanowire before and after sputtering. Solid lines are 
the theoretical fits according to LAMH model\cite{LAMH}. Fitting parameters are listed in table 
\ref{parameters}.} 
\label{Laama}
\end{figure}

\begin{table}
\caption{\label{parameters}The fitting parameters for the LAMH model\cite{LAMH} for the data from 
Fig. \ref{Laama}.}
\begin{ruledtabular}
\begin{tabular}{lccc}
&Original&After 1st sputtering&After 2nd sputtering\\
\hline
$T_{\rm C}$ [K]&1.285 & 1.316 & 1.371\\
$R_{\rm N}$ [$\Omega$]&82&142&380\\
$B_{\rm C}(0)$ [mT]&8.0&7.5&7.0\\
$l$ [nm]&15.8&12.8&9.5\\
$\sqrt{\sigma}$ [nm]&65&55&39\\
\end{tabular}
\end{ruledtabular}
\end{table}

Fig. \ref{Laama} shows a typical effect of sputtering on the shape of superconducting transition 
of a wire. It is clearly seen, that the transition becomes wider and the critical temperature 
$T_{\rm C}$\cite{tc1,tc2} and the normal state resistance $R_{\rm N}$ changes. All these 
observations indicate that the wire cross-section has decreased. 

The dimensionality of a superconductor is determined by the temperature-dependent coherence length 
$\xi(T)$. For the 'dirty limit' superconductors (mean free path $l$ is smaller than $\xi$) the 
effective coherence length $\xi(T) = 0.85 (l \xi_{\rm BCS}(T))^{1/2}$\cite{Tinkham}, where the 
zero-temperature BCS coherence length for aluminum is $\xi_{\rm BCS}(0) \sim 1.6 \ \mu$m. Thus, 
formally, any size superconductor sufficiently close to the critical temperature $T_{\rm C}$ can 
be considered as 'low-dimensional'. However, if to restrict ourselves to temperatures apart from 
the fluctuation range and the reasonably 'dirty' samples ($l \sim$ 10-30 nm), one can consider an 
aluminum wire with the effective diameter $\sqrt{\sigma} <$ 100 nm  ($\sigma$ being the wire 
cross-section) as one-dimensional (1-D).

The shape of the bottom part of a superconducting transition $R(T)$ of not too 
narrow\cite{Zaikin,Golubev} 1-D wires is described by the model of temperature activated phase 
slips\cite{LAMH,Lukens}. The effective resistance exponentially depends on the ratio of the 
condensation energy of a minimum size superconducting domain of size $\sim\sigma\xi$ and the 
thermal energy $k_{\rm B}T$\cite{LAMH}:
\begin{equation}\label{Reff}
R_{\rm eff}(T)\sim R_{\rm N}\frac{L}{\xi}\exp\left(-\frac{B_C^2\Omega}{k_{\rm B}T}\right) ,
\end{equation}
where $R_{\rm N}$ is the normal state resistance, $L$ is the length of the wire, $B_{\rm C}(T)$ is 
the critical magnetic field, and $\Omega=K\xi\sigma$ is the volume of the so-called phase-slip 
center (minimum size of a superconductor to be driven normal). Coefficient $K$ should be of the 
order of  one: $K\sim 1$, relating the geometrical size to the effective one. The complete 
expression for the effective resistance used for $R(T)$ data fitting includes other 
terms\cite{LAMH}, being dependent, for example, on the ratio between the measuring current and the 
critical current. However, the used measuring currents ($\sim$ 10 nA) were much smaller than the 
critical value. Hence, these terms do not contribute quantitatively and are skipped in 
(\ref{Reff}) for simplicity.

The cross-sections $\sigma$ obtained from the measurements of the normal state resistance 
correlate well with the ones used in the fitting procedure (Table \ref{parameters}). The common 
parameters used for the fits are the sample length $L=10\ \mu$m, $\xi_{\rm BCS}(0)=1.6\ \mu$m, 
$K=0.7$, and the product of resistivity and mean free path $\rho l = 4.3\cdot 10^{-16}$ 
$\Omega$m$^2$. The critical temperatures $T_{\rm C}$ used in fitting procedure (Table 
\ref{parameters}) correspond well to the experimentally observed onsets of superconductivity (Fig. 
\ref{Laama}). The increase of the critical temperature with the reduction of the aluminum wire 
cross-section (and, in general, the thickness of a film) is a well-known effect. Commonly accepted 
explanation for this phenomenon is not yet settled, while various models are currently 
discussed\cite{tc1,tc2}. So far, no traces of the quantum phase slip 
phenomena\cite{Zaikin,Golubev,Giordano,Bezryadin} have been detected. At least, down to the 
aluminum wire effective diameter $\sqrt{\sigma} \sim$30 nm.

The fitting of the experimental data with model calculations\cite{LAMH} clearly indicates the 
reduction of the wire cross-section (Table \ref{parameters}) while subsequent sputtering sessions. 
The absence of artefacts on the experimental $R(T)$ dependencies (Fig. \ref{Laama}) proves that 
the dry ion etching does not cause 'serious' damage to the sample (voids or constrictions), but 
removes the material from the surface gently and in a controllable way. Qualitatively similar 
results were obtained on few tens of samples.

\subsection{Sputtered single electron transistors}

A single electron transistor consists of an island isolated from the environment via two tunnel 
junctions, and a gate electrode which is not important for this study. The characteristic 
parameter of a single electron transistor is the charging energy, which is the energy required to 
add one extra electron into the island of an SET, 
\begin{equation}\label{Ec}
E_{\rm C}=\frac{e^2}{2C_{\rm \Sigma}} ,
\end{equation}
where $C_\Sigma =C_1 +C_2 +C_0$ is the sum of the capacitances of the junctions and the 
capacitance of the central island to the ground. Usually it is assumed that the ground capacitance 
$C_0$ is negligible and the capacitances of the junctions are equal, $C_1 =C_2 \equiv C$. The 
charging energy is of particular interest in this study because it is inversely proportional to 
the area of the junctions. 
This can be seen by substituting the expression for the plate capacitor $C=\varepsilon 
\varepsilon_0 A/d$ to Eq. (\ref{Ec}), where $A$ is the junction area, $d$ is the thickness and 
$\varepsilon$ the dielectric constant of an insulating barrier. Thus the charging energy depends 
on the size of the junctions.

\begin{figure}[ht] 
\includegraphics[width=90mm]{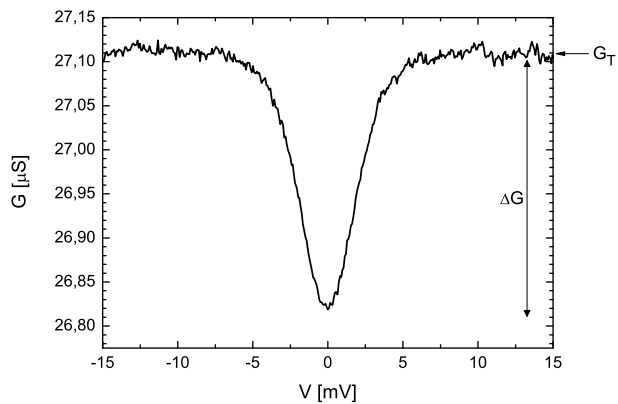}
\caption{Typical conductance dip of a SET at liquid helium temperature 4.2 K.} 
\label{expeak}
\end{figure}

The conductance of a SET as a function of the bias voltage at low temperatures shows a 
dip\cite{CBT}. In this study the parameter of interest is the relative height of the $\Delta G /G$ 
dip (Fig. \ref{expeak}), which is proportional to the charging energy\cite{CBT}:
\begin{equation}\label{cbt1}
\Delta G / G_{\rm T}=\frac{E_{\rm C}}{6k_{\rm B} T}.
\end{equation}
By measuring the charging energy at liquid helium temperature 4.2 K before and after ion beam 
etching, one can determine how the areas of junctions have changed by sputtering. Here a natural 
assumption has been made that the thickness of the insulating layer and its dielectric constant 
$\varepsilon$ are not altered while etching.

\begin{figure}[hbt]
\includegraphics[width=80mm]{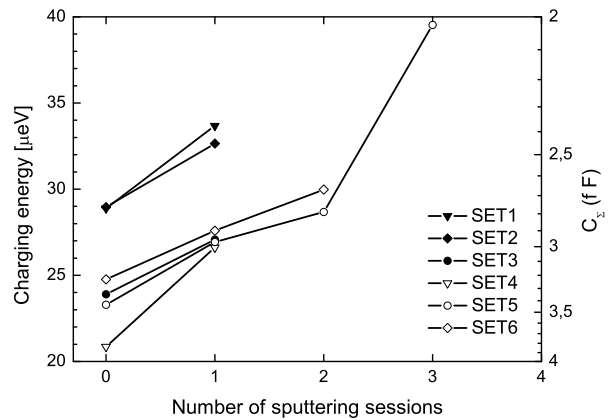}
\caption{Charging energy of single electron transistors (left axis) and the total capacitance 
(rigth axis) as functions of the number of sputtering sessions.} 
\label{nobel}
\end{figure}

Figure \ref{nobel} shows the charging energy of different SET samples measured after fabrication 
and each sputtering session. It is clearly seen that the sputtering increases the charging energy 
gradually, indicating that the method is capable of reducing the areas of the tunnel junctions. 
Similar effect has been already observed\cite{nakamura}. As it follows from equation (\ref{cbt1}), 
the relative height of the conductance dip is inversely proportional to the junction area. 
Therefore, one can conclude from the Figure \ref{nobel}, that the junction areas of the sample 
"SET5" have reduced by $\sim$40 \% with respect to the original after three sputtering sessions. 
The tunnel junction resistances also increased while etching, additionally indicating the 
reduction of the tunnel junction areas. For instance, in sample "SET5" the original resistance was 
37 k$\Omega$, and it became 84 k$\Omega$ after third sputtering. Multisession sputtering was 
performed on many of the samples without damaging them. Actually the sputtering was not always the 
cause of the broken SETs; many of those were destroyed while making the electric contacts.

Since the tunnel junctions are formed by thin oxide layers in between aluminum electrodes, there 
exists a possibility of natural 'aging' of the samples at normal atmospheric conditions leading to 
changes of characteristics. Reference measurements were performed to rule out the possibility that 
charging energy changes 'by itself' by this natural aging. The charging energy of these samples 
was measured repeatedly during the time period of several days. No increases in charging energy 
were observed. So we can be sure that the aging effect is negligible.

In the first sputtering session a surface layer of approximately 25 nm was removed from the 
electrodes and the island forming the SET. In the following 5-7 nm per session were etched. The 
behavior of sample "SET 5" is a bit surprising: although the sputtering conditions and estimated 
thickness of the removed layers were the same in sessions 2 and 3, the increase of the charging 
energy is much larger after the third session. It implies that either the sputtering rate changes 
as the dimensions of the SET structure become smaller, or a certain critical state of the system 
has been reached. 

\begin{figure}[ht] 
\includegraphics[width=80mm]{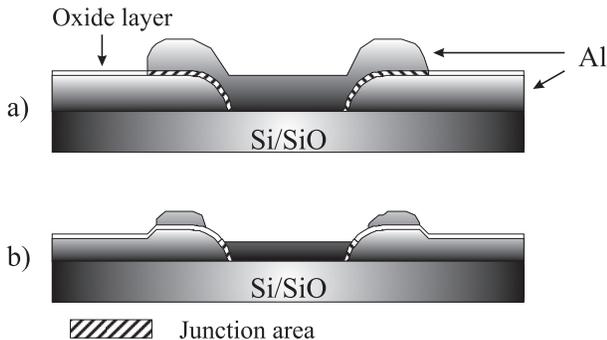}
\caption{Schematic sideview of a single electron transistor. Figure a) represents the SET just 
after fabrication and b) after sputtering.} 
\label{SETside}
\end{figure}

What could this critical state be? Fig. \ref{SETside} a) shows a schematic drawing of the SET 
structure seen from the side just after fabrication. 
Due to of the shadow evaporation method the metal parts do overlap and the oxide layer gets its 
characteristic form having approximately vertical and horizontal parts. While sputtering, one 
finally reaches a state pictured in Fig. \ref{SETside} b). The metallic contact between the metal 
of the island and the metal on top of the electrode is lost and 'suddenly' the effective junction 
area is much smaller. We assume, that the abrupt increase after third sputtering session for the 
structure "SET5" (Fig. \ref{nobel}) is at least partly due to this threshold effect. 

The sputtering rate should not be necessarily the same for macroscopically large and nanosized 
objects. The calibration of sputtering rate in our experiments was done by measuring the height of 
the step developed between sputtered and non-sputtered regions on the large aluminum contact pads. 
Although the profilometer provides high vertical resolution ($\sim$5-7\AA), the lateral dimensions 
of the finest parts in the nanostructures studied were not resolvable. Therefore, if at certain 
stage of sputtering the etching rate of the finest parts has dramatically increased, this would 
result in fast reduction of the areas of tunnel junctions and abrupt increase of the observed 
charging energy, e.g., after third sputtering session (Fig. \ref{nobel}). The possibility of such 
a scenario is currently under investigation. At present, knowledge about interaction of ions with 
low dimensional objects, like nanowires, ultra-small tunnel junctions and nano-islands, is rather 
scarce. Ion sputtering of nanosized objects has not been well explored yet, and various aspects of 
this method still have to be studied and developed in order to achieve a level suitable for 
various applications in nanofabrication.

\begin{figure}[h]
\includegraphics[width=85mm]{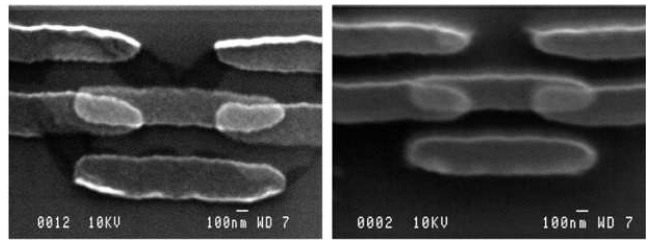}
\caption{SEM images of SET. Left and right images represent the same sample before and after ion 
beam etching, respectively.} 
\label{smooth}
\end{figure}

SEM and AFM imaging of the sputtered samples revealed no strongly developed topography on the 
surface normally attributed to a high fluence ion irradiation. No trenches, craters or other 
extended defects on the surface were observed. On the contrary, the surface of the sputtered 
aluminum structures and their topography became smoother after sputtering when compared to 
as-fabricated state (Fig. \ref{smooth}). It should be noted that single electron transistors are 
usually considered as very fragile to stay 'alive' under experimental manipulations. Nevertheless, 
in our experiments both aluminum nanowires and SETs showed a very high degree of stability under 
high current ion irradiation. Even those SET samples which were 'destroyed' in a sense that the 
resistance became infinite, SEM and AFM observations revealed no breakages or discontinuity. This 
peculiarity of SETs is still unexplained: they may show infinite resistance and still look 
perfect. In our experiments any radiation damage fatal for the performance of the wires and SETs 
should be smaller than $\sim$5 nm in size, otherwise they would be detected by our SEM and AFM. 

\section{Conclusions}

We have demonstrated that the ion beam sputtering can be effectively used for reducing the 
dimensions of prefabricated metallic nanostructures without degradation of their properties and 
performance. The applicability of the method has been verified with aluminum nanowires and SETs. 
In the former case, the diameter of the wires was reduced from the initial 60-70 nm down to 
$\sim$30 nm. When the method was applied to SETs, the charging energy was found to increase 
indicating that the total area of the tunnel junctions decreased. Therefore, it was shown that 
dimensions of the nanostructures can be reduced by ion sputtering in a controllable and 
reproducible way.

By virtue of the surface nature of sputtering phenomenon, the method was proved to be very 
'gentle' in a sense that it allows to decrease the dimensions of delicate nanostructures by 
gradually removing the surface layers without introducing any changes into the interior. 
Reproducibility and controllability provided by the method imply that ion sputtering is in general 
applicable to nanoelectronic components and circuits containing nanosized wires and tunnel 
junctions. The fact that the tunnel junctions 'stay alive' while sputterings makes the range of 
applicability of the method wider. For instance the operational temperatures of single electron 
devices can be extended by increasing the charging energy. It is believed, that bombardment with 
low energy inert argon gas ions causes no chemical reactions. The method is envisaged to be 
applicable to circuits based on any kind of metals, semiconductors, insulators and their 
combinations, regardless of the chemical composition and morphology. Since the method can be 
applied repeatedly to the same sample, gradual reduction of dimensions is achievable in those 
applications where the size effect is studied or employed. Instead of fabricating a number of 
samples of different sizes and comparing their properties and performance, the sputtering method 
allows to work only with a single sample, thereby, avoiding uncertainties due to the 
circumstantial factors in fabrication.

\begin{acknowledgments}
This work has been supported by the Academy of Finland under the Finnish Centre of Excellence 
Programme 2000-2005 (Project No. 44875, Nuclear and Condensed Matter Programme at JYFL) and the EU 
Commission FP-6 NMP-1 "ULTRA-1D" project No:505457-1 "Experinmental and theoretical investication 
of electron transport in ultra-narrow 1-dimensional nanostructures".
\end{acknowledgments}

\end{document}